\documentclass[%
 reprint,
nobibnotes,
 amsmath,amssymb,
 aps,
]{revtex4-1}

\usepackage{graphicx}
\usepackage{dcolumn}
\usepackage{bm}
\usepackage[titletoc,title]{appendix}
\usepackage{amsmath}

\usepackage[
top    = 1.8cm,
bottom = 1.5cm,
left   = 2cm,
right  = 2cm]{geometry}

\usepackage{epstopdf}

\newcounter{mysubtable}
\newcommand\modcounter{%
  \refstepcounter{mysubtable}%
  \renewcommand{\thetable}{A\arabic{table}}%
}

\begin{document}


\title{Assembly of nothing: Equilibrium fluids with designed structured porosity}

\author{Beth A. Lindquist}
\affiliation{McKetta Department of Chemical Engineering, University of Texas at Austin, Austin, Texas 78712, USA}

\author{Ryan B. Jadrich}
\affiliation{McKetta Department of Chemical Engineering, University of Texas at Austin, Austin, Texas 78712, USA}

\author{Thomas M. Truskett}
\email{truskett@che.utexas.edu}
\affiliation{McKetta Department of Chemical Engineering, University of Texas at Austin, Austin, Texas 78712, USA}

\date{\today}

\begin{abstract}

\noindent Controlled micro- to meso-scale porosity is a common materials design goal with possible applications ranging from molecular gas adsorption to particle size selective permeability or solubility. Here, we use inverse methods of statistical mechanics to design an isotropic pair interaction that, in the absence of an external field, assembles particles into an inhomogeneous {\em fluid} matrix surrounding pores of prescribed size ordered in a lattice morphology. The pore size can be tuned via modification of temperature or particle concentration. Moreover, modulating density reveals a rich series of microphase-separated morphologies including pore- or particle-based lattices, pore- or particle-based columns, and bicontinuous or lamellar structures. Sensitivity of pore assembly to the form of the designed interaction potential is explored.

\end{abstract}

\pacs{Valid PACS appear here}
\maketitle





Synthesis of materials with controlled micro- to meso-scale porosity is a design challenge ripe for new, innovative approaches. For example, ordered porous zeolite solids--commonplace for their use as molecular sponges to absorb gases--have recently inspired an extension to the liquid state via molecular cages dispersed in a solvent of larger excluded molecules to form an absorbing ``pore fluid''~\cite{pore_liquid_cages}. At the colloidal level, control over the effective particle interactions can lead to equilibrium structures with large percolating voids (i.e., connected regions of solvent-occupied, but particle-free, space), as was recently highlighted via a computational study of indented particles with depletion attractions~\cite{pore_liquid_ind_colloids}. Prior theoretical work~\cite{postulated_phases_sear,mean_field_assembly_1,mean_field_assembly_2,simulated_phases} suggests that even a single-component system interacting via an isotropic pair potential comprising competitive attractive and repulsive components can form an equilibrium fluid with a porous structure. Such systems also appear to have modified vapor-liquid phase behavior leading to coexistence involving modulated phases~\cite{simulated_phases}. Even a simple two-dimensional lattice model with competing interactions was shown to assemble into a fluid phase with ``bubbles'' or pores, among other microphases~\cite{lattice_bubbles_2}; a two-dimensional continuum theory yields similar predictions~\cite{2D_microphase}. Such results are reminiscent of experimentally observed interfacial structures of colloids with competing interactions~\cite{jaime_expt}. Here, we explore the prospect of using inverse methods of statistical mechanics to discover which (if any) isotropic pairwise interactions spontaneously assemble particles into an inhomogeneous three-dimensional fluid matrix surrounding ordered, monodisperse spherical pores of prescribed size. We further study how the microphase structure of such a model material would respond to changes in temperature and particle concentration or depend on the specific form of the interaction. The ability to control the size and spatial organization of pore structures in fluids, via interactions and thermodynamic parameters, could help pave the way for novel colloidally assembled structures with programmable particle-size-selective permeability/solubility similar to materials templated via block-copolymer assembly~\cite{block_copolymer_membranes} (with selective etching of one block component to create void space).  

Modern computational coarse-graining~\cite{IBI_and_RE_ID,RE_ID} and inverse design strategies are ideally suited for the optimization of pair interactions to realize complex, self-assembled material architectures. Previously, inverse design has led to the discovery of isotropic, convex-repulsive pair interactions that stabilize exotic, open crystalline arrangements (e.g., honeycomb and diamond lattices)~\cite{ID_crystals_TS,ID_crystals}. Another example is the design of pair interactions that promote the formation of a fluid of well-defined equilibrium particle clusters (i.e., microscopic liquid droplets)~\cite{ID_clusters}. Here, we use iterative Boltzmann inversion (IBI)~\cite{IBI_and_RE_ID,ID_clusters} to find the dimensionless, isotropic pair potential \(\beta u_{\text{IBI}}(r)\) [where $\beta=(k_{\text B}T)^{-1}$, $k_{\text B}$ is the Boltzmann constant, and $T$ is temperature]  which best reproduces, at equilibrium, the radial distribution function (RDF), \(g_{\text{tgt}}(r)\), of our target structure: a fluid matrix of particles surrounding a lattice of pores or voids with prescribed size. 

\begin{figure}
  \includegraphics{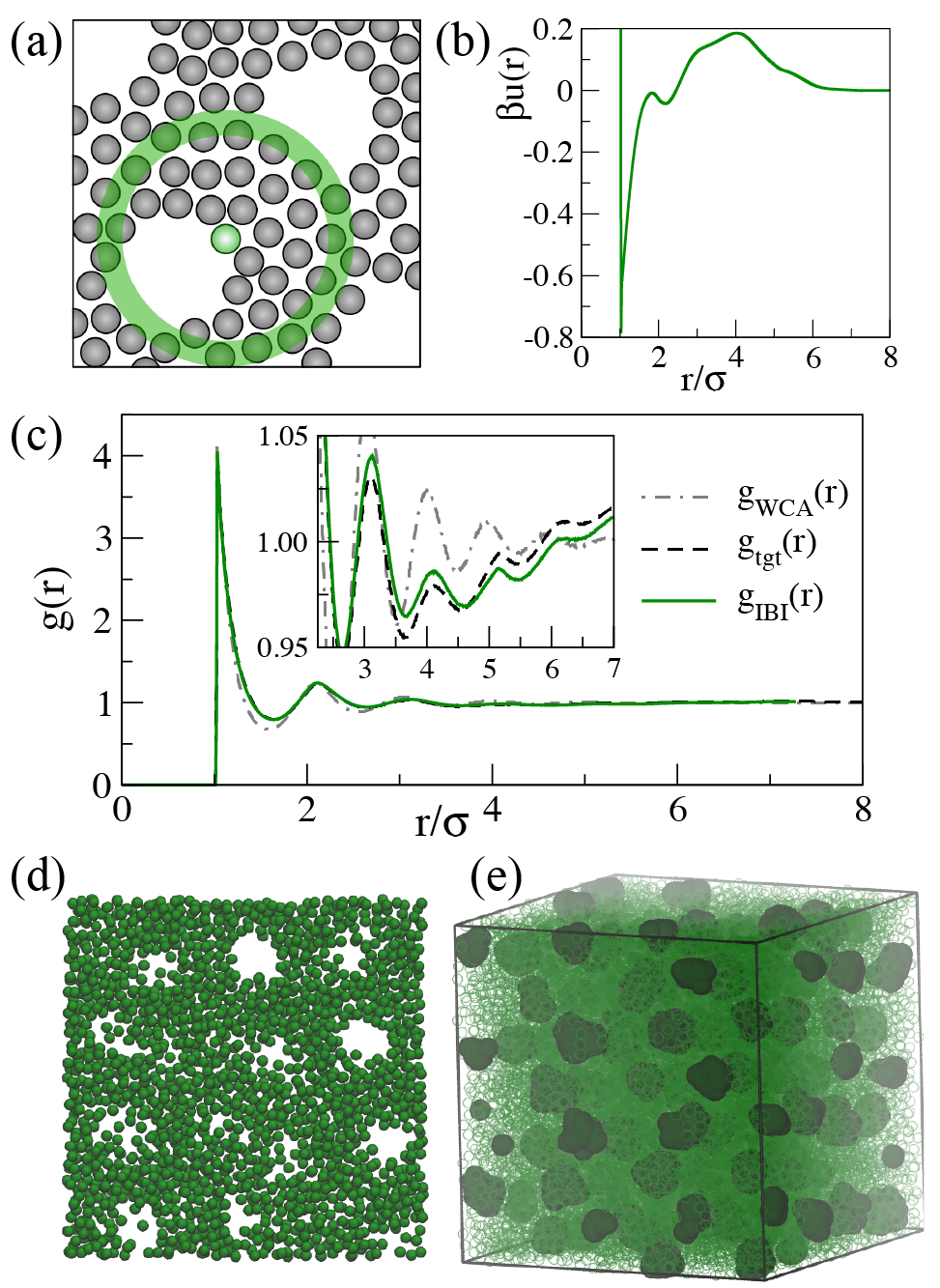}
  \caption{Design of a pair potential, \(\beta u_{\text{IBI}}(r)\), that assembles particles into a fluid matrix surrounding a lattice of pores. (a) Schematic for a system with matching pore diameter and pore wall thickness, where the repulsive hump in \(\beta u_{\text{IBI}}(r)\) is shown in green for the highlighted particle. (b) The optimized pair potential, \(\beta u_{\text{IBI}}(r)\). (c) Radial distribution functions for the targeted configurational ensemble of the fluid matrix particles (\(\eta=0.31\)), a system of particles with the IBI-optimized pair potential (\(\eta=0.31\)), and an isotropic, single-component WCA fluid with the same isothermal compressibility (\(\eta=0.42\)). (inset) Zoomed view of main frame. (d) Image of a \(3\sigma\) thick slab extracted from an equilibrium simulation configuration of particles with the IBI-optimized pair potential, illustrating pores. (e) Dark green regions show a three-dimensional visualization of the assembled BCC pore structure and represent the union volume of all \(2\sigma\) diameter test spheres that could be successfully accommodated into the configuration without overlapping lighter green fluid particles interacting via the IBI-optimized pair potential.}
  \label{fgr:FigureRDF}
\end{figure}

The configurational ensemble of the target structure is generated from a molecular simulation of an inhomogeneous fluid of effective hard-core Weeks-Chandler-Andersen (WCA) particles of diameter $\sigma$ embedded in the interstitial space of a lattice of larger WCA particles of diameter 4$\sigma$; the latter enforces the ordered pore structure by excluding the target fluid particles. 
The large particles were fixed on an FCC lattice with a nearest-neighbor distance of 7.4\(\sigma\). This choice was motivated to balance two competing effects: (1) to maximize the pore signature in \(g_{\text{tgt}}(r)\) while (2) furnishing pore ``walls'' formed by the fluid matrix with a thickness comparable to the pore size (i.e., a pore center-to-center distance of approximately two times the pore diameter). The latter consideration was inspired by preliminary optimizations for several target simulations, which showed that the potential developed a repulsive hump centered near the pore size (irrespective of whether that potential successfully formed pores). This occurs so that a particle bordering a pore will impede other particles from entering the pore on the opposite side, as shown on the leftmost pore in the scheme in Fig. ~\ref{fgr:FigureRDF}a, where the repulsive hump for a selected particle is highlighted in green and the schematic is appropriately scaled. A system where the pore wall width and pore diameter are matched is optimal for description with a pair potential so that the highlighted particle can also contribute to the repulsive barrier at the edge of an adjacent pore, as shown for the rightmost pore in Fig. ~\ref{fgr:FigureRDF}a.
Additional details regarding the construction of the target system, including the effect of varying the large particle template, can be found in the Simulation and Analysis Details Section of the Appendix.

\begin{figure}
  \includegraphics{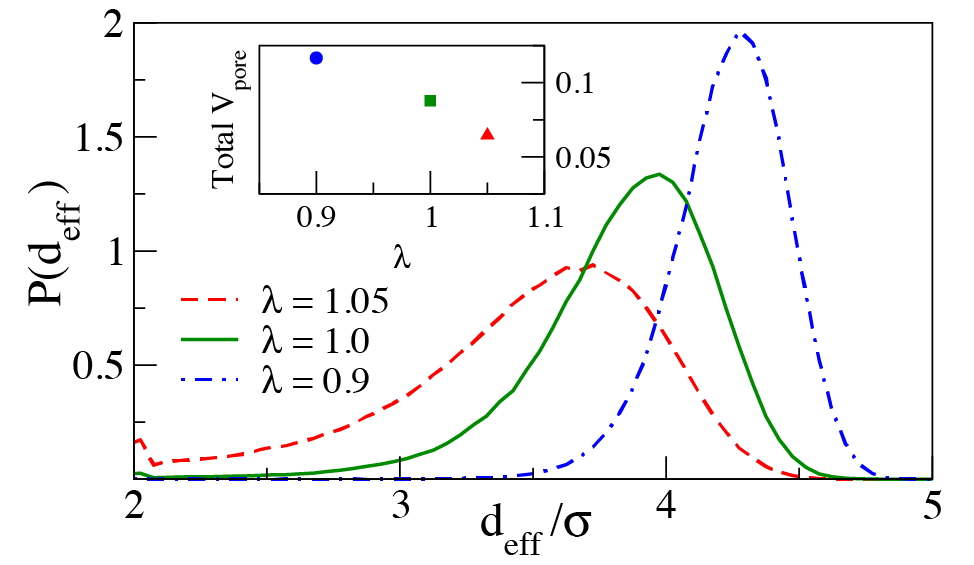}
  \caption{Pore size probability distribution function \(P(d_{\text{eff}})\) for different choices of the temperature rescale parameter \(\lambda\) corresponding to the change \(T \rightarrow \lambda T\) or equivalently \(\beta u_{\text{IBI}}(r) \rightarrow \beta u_{\text{IBI}}(r)/\lambda\). 
  (inset) Total volume of the pores as a function of \(\lambda\).}
  \label{fgr:Temp}
\end{figure}

\begin{figure*}[ht!]
  \includegraphics{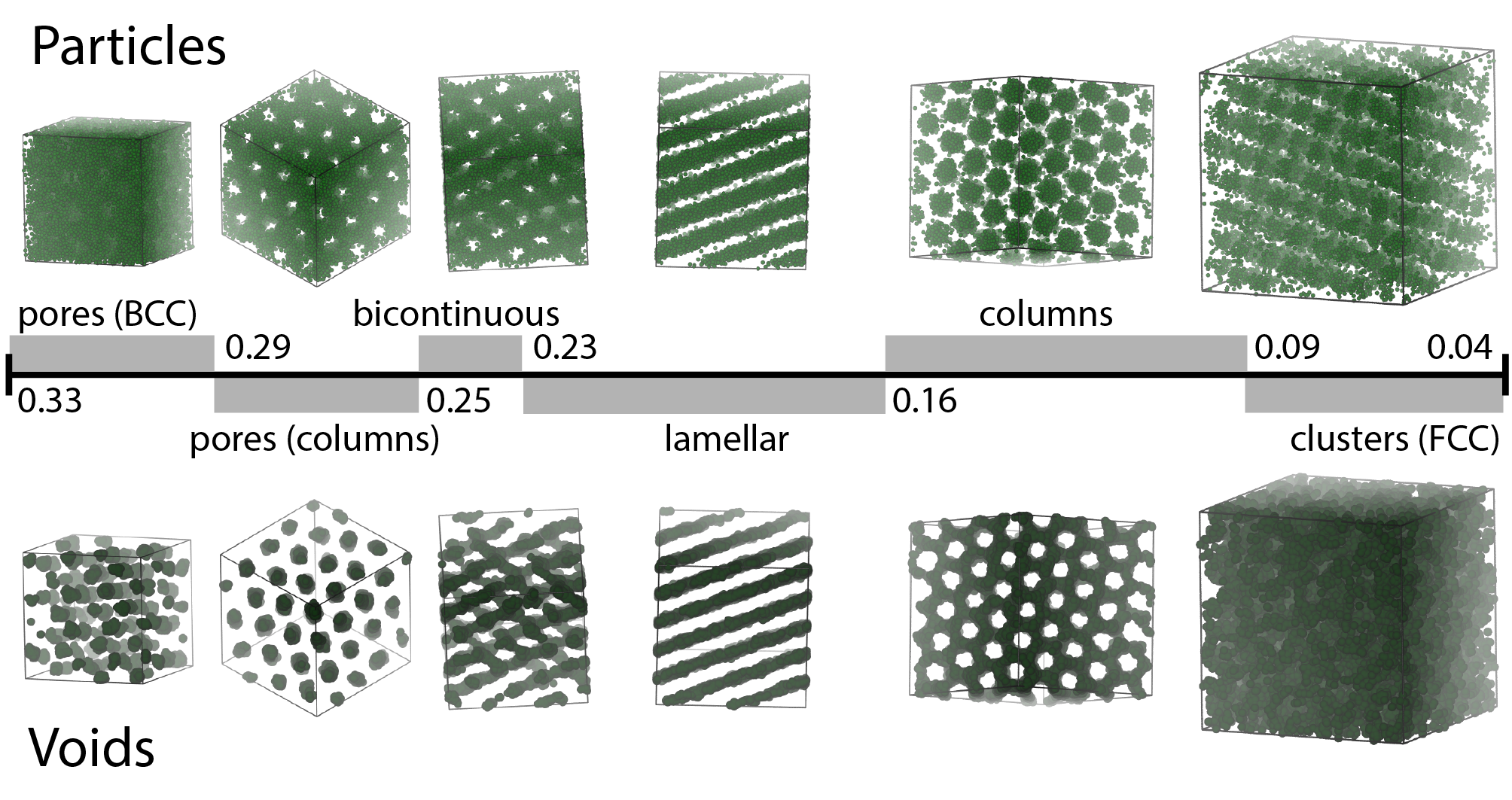}
  \caption{Microphase diagram of the IBI-designed model as a function of $\eta$, with representative snapshots of each phase. Particles are depicted on the top row, and the inserted void particles are shown on the bottom row.}
  \label{fgr:PhaseDiagram}
\end{figure*}

The IBI-optimized pair potential \(\beta u_{\text{IBI}}(r)\) is shown in Fig.~\ref{fgr:FigureRDF}b. Even though there may be non-trivial many-body effects in an actual pore-forming system that are not captured by this strategy, an understanding of an isotropic pair potential that forms pores is useful because it 1) guarantees that many-body potentials exist that will also generate pores, and 2) reveals what those many-body interactions must map onto if they are projected into a pair interaction form. From prior work on microphase-separated states, it is not surprising that \(\beta u_{\text{IBI}}(r)\) possesses competitive attractions and repulsions. The general form of the potential is similar in spirit to the cluster fluid-forming potentials that we reported in previous work~\cite{ID_clusters}, though here the attractive well is generally narrower and deeper and the repulsive hump is broader but shorter when compared to clusters of similar size to the pores studied here. Common to both potentials is that the relevant lengthscales of the microphase-separated objects are directly encoded by the potential: for clusters, the position of the repulsive hump was directly related to cluster size, and here the repulsive hump is maximal at the prescribed pore size. 

Also similarly to our previous cluster work, we observed good convergence in the IBI scheme. 
As can be seen in Fig.~\ref{fgr:FigureRDF}c, \(\beta u_{\text{IBI}}(r)\) is indeed able to closely reproduce the equilibrium RDF of the inhomogeneous, target matrix of fluid particles at a packing fraction, \(\eta=\pi \sigma^3N/6V = 0.31\) where \(N\) is the number of particles and \(V\) is the volume. For comparison, the RDF of an isotropic, single-component WCA fluid with the same isothermal compressibility (\(\eta=0.42\)) is also shown. Interestingly, the RDF of a dense fluid matrix surrounding pores (whether from the target or IBI model structure) is only subtly different from that of the equi-compressible isotropic WCA fluid, with the former being distinguished from the latter by its relatively suppressed correlations on the scale of the pore size, as highlighted in the inset. However, despite its success at reproducing the pair correlations of the target structure, it is not obvious that a system of particles interacting via \(\beta u_{\text{IBI}}(r)\) will also necessarily capture the desired pore structure of the target ensemble, which depends on many-particle static correlations.     

Nonetheless, we do find that pore structures very similar to those templated by the ordered lattice of large WCA spheres of the target system emerge with the IBI-optimized interaction, \(\beta u_{\text{IBI}}(r)\). Hints of the pore morphology can be seen in Fig.~\ref{fgr:FigureRDF}d from a visual representation of the particles within a 3\(\sigma\) thick slab obtained from an equilibrium configuration. However, for a more quantitative characterization, `pores' defined via a clustering analysis~\cite{CSD_1} as union volumes of all \(2\sigma\) diameter test spheres that could be successfully accommodated into the configuration without overlapping the fluid matrix particles were also examined. As described in the Appendix, the size of an individual pore in a configuration is computed from its volume, obtained by Monte Carlo integration of overlapping inserted test particles, and is reported here as the diameter $d_{\text{eff}}$ of an equivolume spherical pore. Fig.~\ref{fgr:FigureRDF}e shows that the pores (compact, dark green entities embedded in the lighter green matrix of fluid particles) actually form a BCC lattice with a lattice constant of \(8.8\sigma\) (i.e., nearest neighbor distance of \(7.6\sigma\)). As can be seen from the solid green line in Fig.~\ref{fgr:Temp}, the most likely effective diameter (\(d_{\text{eff}}\)) of an individual pore is \(3.97\sigma\), very close to desired \(4\sigma\) pore size of the target ensemble. In other words, the IBI pair potential optimized to precisely match the RDF of the target ensemble also faithfully reproduces the ordered pore morphology of the target structure.   

We next evaluate the sensitivity of the pore morphology of the system with the IBI-optimized pair potential to rescaling the temperature by a factor \(\lambda\) as \(T \rightarrow \lambda T\) [equivalent to rescaling \(\beta u_{\text{IBI}}(r) \rightarrow \beta u_{\text{IBI}}(r)/\lambda\)]. Fig.~\ref{fgr:Temp} shows the pore size distribution function for equilibrium simulations with both mild cooling (\(\lambda=0.9\)) and heating (\(\lambda=1.05\)). With increasing temperature, we note the pores monotonically shrink and have larger size fluctuations, with no evident porosity accessible to the \(2\sigma\) test particles by \(\lambda = 1.1\). The inset in Fig.~\ref{fgr:Temp} shows that the total volume of the simulation box taken up by the pores correspondingly decreases with increasing temperature. The opposite trends occur with cooling--pores grow and become more uniform in size--until eventually the pores condense into ``columns'' of void space (by \(\lambda = 0.8\)). Nonetheless, bracketed by these two extremes, there is a range of temperature where approximately spherical pores form in a BCC lattice, and their size can be tuned via modification of the temperature. 

Although small changes in packing fraction ($\eta$) also allow for tuning the pore size, we find that larger changes in $\eta$ give rise to a rich diagram of microphase-separated void-particle morphologies. Fig.~\ref{fgr:PhaseDiagram} shows configuration snapshots of the various microphases associated with \(\beta u_{\text{IBI}}(r)\) (highlighting particle and void structures, respectively) along with the corresponding density range for each phase. Densities of intermediate character are assigned to the phase that they most greatly resemble, with some buckling and/or defects present away from the optimal density for each phase. Upon reducing particle concentration from the optimized pore lattice structure, the pores first coalesce into void columns. Eventually the porous columns give way to a bicontinuous phase, where the void space and particles are interpenetrating. This gyroid-like phase then flattens out into lamellar sheets of alternating particles and voids. Lamellar sheets have been predicted for a single-component model with competing attractions and repulsions that also displays clustering behavior, but the possibility of pore lattices was not explored~\cite{DFT_cluster_lamellar}. In two dimensions, an analogous striped phase has been observed in addition to a clustered phase for various models~\cite{lattice_bubbles_2}.  

The above phase progression of the void space has been seen in diblock copolymers as the relative amount of each block is tuned~\cite{block_coplymers}. Here, $\eta$ controls the ratio of void and particle-filled space. Upon further expansion from the lamellar phase, columns of particles are then formed, followed by clusters of particles as shown on the right hand side of Fig.~\ref{fgr:PhaseDiagram}. The particle clusters form with a preferred size, with radii of gyration ranging from \(R_{g}=2.00\sigma\) to \(2.19\sigma\) as particle concentration is decreased; see the Appendix. The low-density region of the microphase diagram illustrates a symmetry between clusters and pores first postulated by Sear and Gelbart~\cite{postulated_phases_sear} and later predicted from theoretical calculations~\cite{mean_field_assembly_1,mean_field_assembly_2}. Intriguingly, microphases containing a single cluster and a single pore were observed from Grand Canonical Monte Carlo calculations of an SALR potential at different densities by Archer and Wilding~\cite{simulated_phases}. This potential also showed a similar progression of phase behavior between these two state points, though the assembled structures had features on the order the size of the simulation box such that only a single column or sheet could form. As a result, it is unclear to what extent finite size effects impacted the characteristics of these phases. Here, we circumvent this difficulty by leveraging inverse design to systematically imbue a prescribed length scale to the features arising from microphase separation. 

Finally, we explored the sensitivity of pore formation to the details of the interactions.  Specifically, we performed new optimizations of the target ensemble using the relative entropy approach~\cite{IBI_and_RE_ID,RE_ID}, restricting the designed pair potential to a variety of functional forms, all containing a single attractive well plus a longer-ranged repulsive hump controlled by four scalar parameters (two energies and two length scales). The analytic forms of the optimized potentials are given in the Appendix. Fig.~\ref{fgr:potentials}a compares \(\beta u_{\text{IBI}}(r)\) to three such optimized potentials, and Fig.~\ref{fgr:potentials}b shows the resulting pore-size PDFs. All potentials form pore lattices, albeit with somewhat reduced pore sizes relative to \(\beta u_{\text{IBI}}(r)\)--though, as we showed above, the pore size can also be modulated via temperature. See the Appendix for corresponding simulation snapshots, where the discrete character of the pores is visually evident.

The variation in the potentials indicates that there are many ways to balance attractions and repulsions in order to achieve a porous microphase separated state, though the repulsive hump tends to be centered near the targeted pore size. Generally, we find that longer-ranged attractions than are traditionally found in microphase-separated models are preferred for forming size-specific pores. Although it might be difficult to realize such effective interactions for micron-sized colloidal particles, it should be possible for nanoscale particles. As an example, potentials of mean force observed in simulations of ligand-coated, charge-stabilized gold nanoparticles display a similar qualitative form~\cite{exp_pore_cluster_system}.
Moreover, in future work, a similar relative entropy approach could be used to limit the range of the attraction for application to colloids as well as to explore the range of parameters that can be employed in the target simulation to successfully make pores, particularly with respect to pore size and pore density. 

\begin{figure}
  \includegraphics{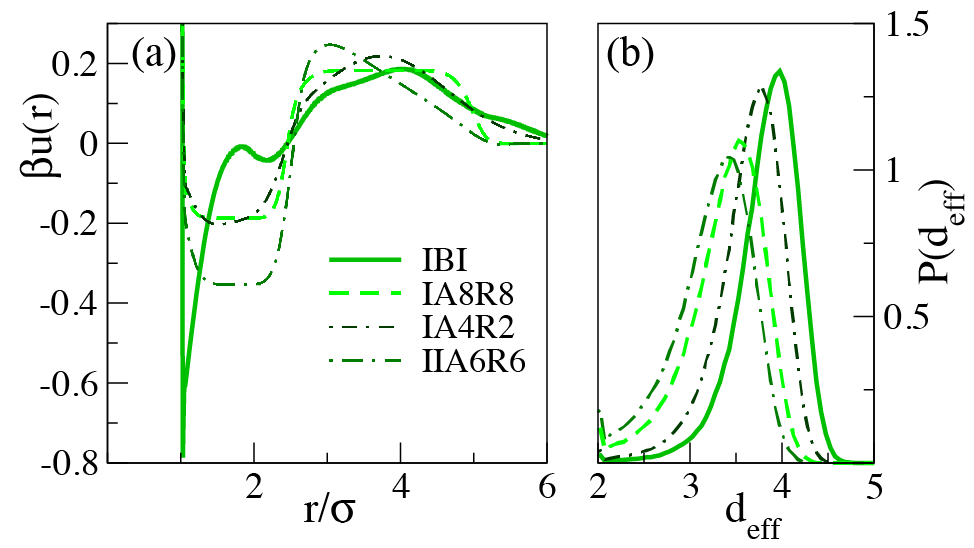}
  \caption{Inverse designed pair potentials that form pores. (a) Potentials optimized with IBI and relative entropy approaches, where the latter are constrained to have a specific functional form (given in the Appendix). (b) Pore-size PDFs of the optimized potentials in (a).}
  \label{fgr:potentials}
\end{figure}

In conclusion, we have used inverse design to discover a pair potential that assembles particles into a fluid matrix surrounding a lattice of pores with prescribed size, in addition to other complex fluid-pore microphases analogous to those seen in diblock copolymer systems (clusters, columns, lamellar sheets, and a bicontinuous phase). We have further demonstrated that the assembly of such microphases can be achieved via a variety of different interaction potentials displaying competitive attractive and repulsive interactions.



\begin{acknowledgments}
This work was partially supported by the National Science Foundation (1247945) and the Welch Foundation (F-1696). We acknowledge the Texas Advanced Computing Center (TACC) at The University of Texas at Austin for providing HPC resources.
\end{acknowledgments}


\setcounter{figure}{0}
\renewcommand\thefigure{A\arabic{figure}}
\renewcommand{\thesection}{\thepart .\arabic{section}}
\renewcommand\theequation{A\arabic{equation}}

\section*{APPENDIX}

\subsection*{Simulation and Analysis Details}
To calculate \(g_{\text{tgt}}(r)\), we simulated $N=5475$ small (S) particles in a frozen matrix of 32 large (L) template spheres to create pores. Small particles are excluded from one another and from large particles via effective hard core repulsions of diameter \(\sigma\) and cross diameter \(2.5\sigma\), respectively. The packing fraction \(\eta=\pi \sigma^3N/6V\) of the small spheres in volume $V$ is 0.31. Simulations were performed using the GROMACS 4.6.5 molecular dynamics package~\cite{GROMACS_1,GROMACS_2} with the 50-25 Weeks-Chandler-Andersen (WCA) effective hardcore potential~\cite{HansenMcDonald} 
\begin{equation} \label{eqn:wca_potential}
\begin{split}
&u_{\text{WCA}}^{(i,j)}(r)\equiv H(2^{1/25}\sigma_{i,j}-r)\\ &\times\Bigg(4 \epsilon \bigg[\bigg(\dfrac{\sigma_{i,j}}{r}\bigg)^{50}-\bigg(\dfrac{\sigma_{i,j}}{r}\bigg)^{25}\bigg]+\epsilon\Bigg)
\end{split}
\end{equation}
where \(\sigma_{S,S}=\sigma\), \(\sigma_{S,L}=2.5\sigma\) and \(H(x)\) is the Heaviside step function. In standard GROMACS units (see GROMACS manual) we set \(\sigma = 1\), \(\epsilon = 1.0\), \(m=30\) (mass), \(dt=0.001\) (time step) and \(T=300\) via velocity-rescale thermostating with characteristic time constant \(\tau=100dt\) and rescaling  every \(10dt\). These conditions yield the dimensionless relations \(\beta \epsilon \approx 0.4\) for the WCA potential and \(dt/t_{0}\approx0.0003\) where \(t_{0}\equiv \sqrt{\sigma^2m/(k_{B}T)}\) is the elementary ballistic time of the molecular dynamics simulation. The choice \(\beta \epsilon\) is not unique; it served only to furnish a steep repulsion mimicking a hardcore--even a totally different functional form that is sufficiently steep would suffice. With this, a target RDF with pores of a prescribed size was generated for the inverse design (ID) step. (Note: in practice, the ID step employs a fixed simulation temperature; however, this is totally arbitrary, and variation would lead to trivial rescaling of the optimized potential such that its thermally non-dimensionalized form is unique).

To check for finite size effects, simulations of all optimized potentials were performed with 18478 particles, which corresponds to increasing the box size used in the ID optimization by a factor of 1.5. Starting from a disordered configuration, a lattice of pores is typically formed within \(\approx1,000,000dt/t_{0}=300\) short time non-dimensionalized units--a relatively short timescale for molecular dynamics simulations. As a zeroth order comparison to a system with solvent, we assume our molecular dynamics ballistic short time unit can be replaced by a short time diffusion scale in a solvent based system. (This also neglects possible complications from hydrodynamic effects.) For \(25nm\), \(100nm\) and \(2\mu m\) particles in common solvents, the estimates for \(t_{0}\) are 0.15ms, 0.01s and 30s, respectively~\cite{time_scale}. Thus, the corresponding times to assembly under our very simplified assumptions are 0.045s, 3s and 150min. While these estimates may seem rather fast, they provide a rough estimate of the expected order for the time to assembly.

As described in the main text, pores were characterized via insertion of spheres with a diameter of \(2\sigma\) such that they did not overlap with the system particles. The pore volumes were determined via Monte Carlo integration of the inserted void particles, where the inserted spheres were assigned to a given pore based on a clustering analysis, where overlapping spheres are taken to be neighbors~\cite{CSD_1}. The resulting volume is then converted to the diameter of a sphere of equal volume. This strategy allows for an evaluation of moderately aspherical pores. Though visual inspection reveals that the pores are relatively spherical, instantaneous fluctuations of the pore walls as well as infrequent particle motions through the pores render the assumption of perfectly spherical pores (often employed in the calculation of pore size distributions) inaccurate for this case, under-predicting the pore size.

As discussed in the main text, the target simulation was chosen such that the pore diameter was similar to the thickness of pore ``walls'' to find a target well suited to a description with a pair potential: the large particles were fixed on an FCC lattice with a nearest neighbor distance of 7.4$\sigma$. We also tried an optimization with a template composed of frozen disordered large particles, the positions of which were generated from a simulation of only the large particles with effective hard core repulsions of diameter 6$\sigma$. All other parameters were identical to those described above. (In the binary target simulation, the large particles still interacted with the small particles via a 4$\sigma$ effective hard core.) We performed 10 target simulations, each using a different simulation snapshot for the template, and found that \(g_{\text{tgt}}(r)\) was essentially invariant with respect to the different frozen templates. We then performed the IBI optimization with the \(g_{\text{tgt}}(r)\) that was averaged over the different frozen templates. Despite the larger mismatch between pore diameter and pore wall size, this potential still yielded discrete pores, positioned on a defective FCC lattice where the pores have significantly more translational motion about their lattice sites. However, the pores are somewhat smaller and thus do not reproduce the targeted pore size as well (the maximum in the pore size distribution function is \(\approx 3.6\sigma\)). Therefore, while there is a range of acceptable templates that can form pores, a target system with a matching pore wall thickness and pore diameter seems to be more accurately reproduced by a pair potential. 

\subsection*{Inverse Design Methodology}
\begin{figure}
  \includegraphics{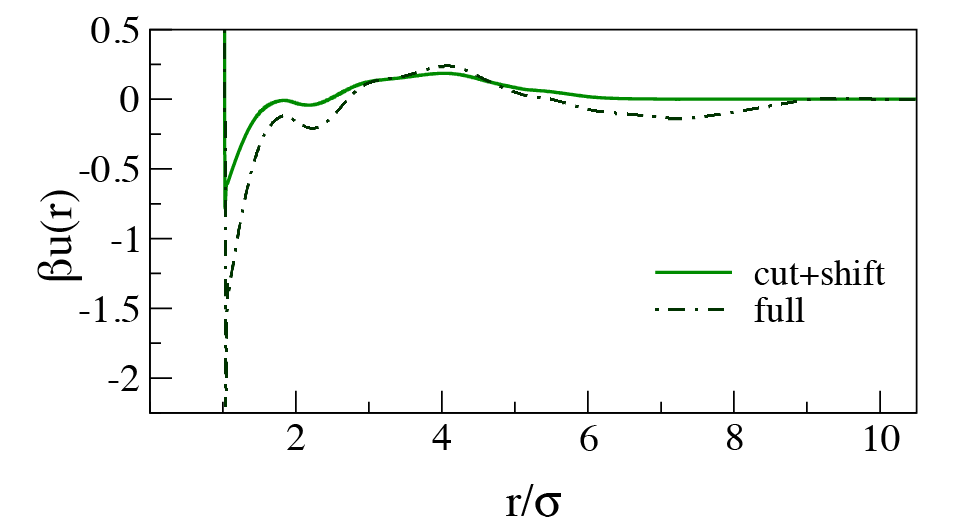}
  \caption{IBI-optimized potentials with \(R_{\text{C}}=10.5\sigma\) and \(7.3\sigma\).}
  \label{fgr:FigureLRPOT}
  
\end{figure}

Unconstrained potential optimizations were performed using iterative Boltzmann inversion (IBI), a powerful inverse statistical mechanics tool from the coarse-graining community~\cite{IBI_and_RE_ID}. IBI assumes nothing about the functional form of the potential and seeks to exactly replicate \(g_{\text{tgt}}(r)\). In practice, the potential is finely discretized, cut and shifted at a finite cutoff distance (\(R_{\text{C}}\)), and spline fitted. In a preliminary IBI optimization, \(R_{\text{C}}\) was set equal to one-half of the box length (\(10.5\sigma\)). This potential, shown in Fig.~\ref{fgr:FigureLRPOT}, formed pores on an FCC lattice; however, when the potential was simulated in a larger box, the system phase separated into a liquid and a gas (though a metastable porous phase was observed en route to macrophase separation). Suspecting that the long-range attraction was responsible for these finite-size effects, and noting from our earlier work that only a single attractive well and repulsive hump were needed to form clusters~\cite{ID_clusters}, we cut and shifted the potential at the minimum at \(7.3\sigma\). We then re-optimized the potential with \(R_{\text{C}}=7.3\sigma\) to yield the final IBI potential. The behavior of the resulting potential was robust with respect to increasing the box size. IBI calculations were performed using the the Versatile Object-oriented Toolkit for Coarse-graining Applications (VOTCA)~\cite{VOTCA_1,VOTCA_2}, which interfaces with the GROMACS 4.6.5 molecular dynamics (MD) package~\cite{GROMACS_1,GROMACS_2}.

For optimizing constrained analytical forms, we utilized Relative Entropy (RE) optimization~\cite{IBI_and_RE_ID,RE_ID,test_of_ID_schemes}. Two functional forms (denoted by the roman numeral indices I and II) for the interactions external to the WCA core are considered
\begin{equation} \label{eqn:re_potential}
\begin{split}
&u_{\text{RE}}^{(i)}(r) \equiv -\epsilon_{1} \textup{exp}\bigg[-\bigg(\dfrac{r-d-\alpha_{1}/2}{\alpha_{1}/2}\bigg)^{n_{1}}\bigg] \\ &+  \epsilon_{2}q_{i}(\alpha_{1},\alpha_{2}) \textup{exp}\bigg[-\bigg(\dfrac{r-d-\alpha_{2}/2-\alpha_{1}}{\alpha_{2}/2}\bigg)^{n_{2}}\bigg]
\end{split}
\end{equation}
where 
\begin{equation} \label{eqn:re_potential_2}
q_{i}(\alpha_{1},\alpha_{2})\equiv \delta_{i,\text{I}} + \delta_{i,\text{II}}\bigg[ \dfrac{-r + d+\alpha_{1}+\alpha_{2}}{\alpha_{2}} \bigg]
\end{equation}
\(\delta_{i,j}\) is the Kronecker delta, [\(\epsilon_{1},\epsilon_{2}\)] are the [attractive, repulsive] strengths, and [\(\alpha_{1},\alpha_{2}\)] are the corresponding ranges. The term for the type II potential effectively changes the symmetric repulsive hump to a linear ramp. Two exponents also appear, [\(n_{1},n_{2}\)] which we set to various numbers to test the flexibility of such potentials to make pores. We use the compact notation, \(i\text{A}n_{1}\text{R}n_{2}\), to denote the type and the exponents as in Figure 4a of the main text. Attraction strengths and ranges are then optimized within the relative entropy framework by maximizing the log-likelihood of sampling the target configurations using a simple gradient descent scheme~\cite{barber_book}. Optimized parameters are provided in Table S1. 

\begin{table}[ht]
\modcounter
\label{tbl:re_parameters}
\centering
\begin{tabular}{  >{\centering\arraybackslash}m{0.95cm}   >{\centering\arraybackslash}m{0.95cm}   >{\centering\arraybackslash}m{0.95cm} |  >{\centering\arraybackslash}m{0.95cm}  >{\centering\arraybackslash}m{0.95cm}  >{\centering\arraybackslash}m{0.95cm}  >{\centering\arraybackslash}m{0.95cm} } 
\hline
Type & \(n_{1}\) & \(n_{2}\) & \(\beta \epsilon_{1}\) & \(\alpha_{1}\) & \(\beta \epsilon_{2}\) & \(\alpha_{2}\) \\
\hline
I & \(4\) & \(2\) & \(0.215\) & \(1.46\) & \(0.218\) & \(2.55\) \\
I & \(8\) & \(8\) & \(0.187\) & \(1.46\) & \(0.183\) & \(2.56\) \\
II & \(6\) & \(6\) & \(0.353\) & \(1.54\) & \(0.322\) & \(2.72\) \\
\hline
\end{tabular}
\caption{Optimal parameters for the constrained potentials.}
\end{table}

\subsection*{Cluster Size Distributions}
\begin{figure}
  \includegraphics{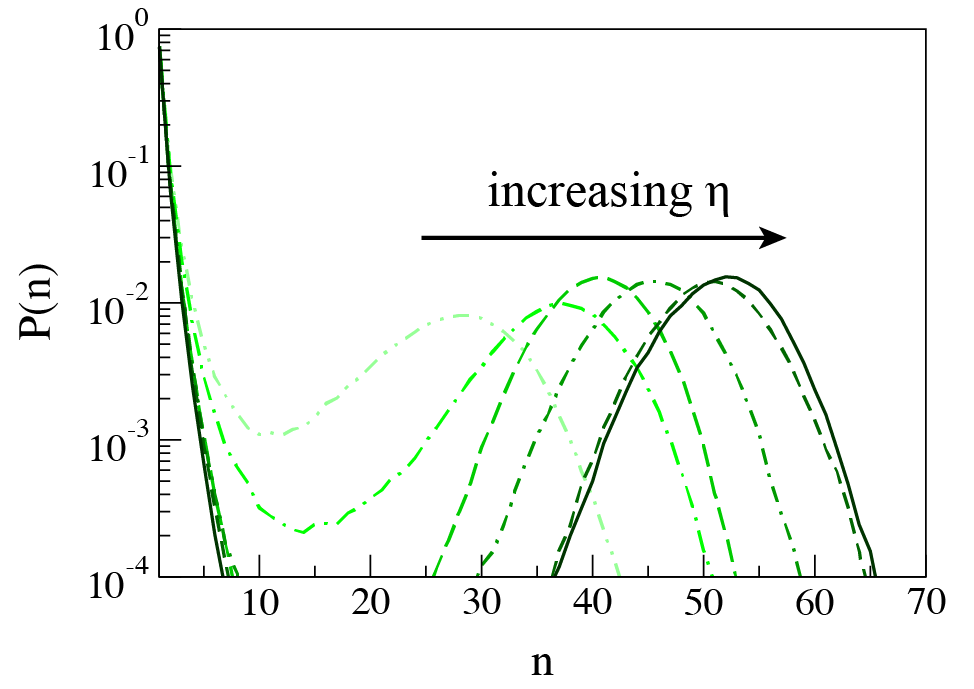}
  \caption{CSDs for \(\eta=0.052\) to \(0.082\) in increments of \(0.006\).}
  \label{fgr:FigureCSD}
  
\end{figure}

Cluster size distributions (CSDs)~\cite{CSD_1} for the clustered phase as a function of density are shown in Fig.~\ref{fgr:FigureCSD}. The CSDs are somewhat sensitive to the choice of distance cutoff, \(R_{\text{CSD}}\), used to determine whether particles are neighbors, though the overall trends remain the same regardless of \(R_{\text{CSD}}\). The relatively broad attractive well in the potential (Fig.~\ref{fgr:FigureLRPOT}) makes determination of \(R_{\text{CSD}}\) non-unique; therefore, a value of \(R_{\text{CSD}}=1.33\sigma\) was chosen such that two discrete clusters were assigned to a single cluster over an order of magnitude less frequently than a single cluster was identified; these dimers are not shown in Fig.~\ref{fgr:FigureCSD} for clarity. Fig.~\ref{fgr:FigureCSD} indicates that, while there is a significant fraction of monomer present, the clusters are reasonably size specific, particularly at the higher densities where there is essentially perfect separation of monomer/small aggregates and clusters. As \(\eta\) is further decreased, the clusters break up and become less size specific, until at approximately \(\eta<0.04\) there is no shoulder in the CSD, indicating no preferred size scale for clustering.

In our prior work~\cite{ID_clusters}, we used IBI to discover potentials which generated a dilute fluid of clusters. By contrast, the clusters in the present study appear on lattice positions as can be seen in Fig. 3 of the main text. Moreover, we visually observe greater particle exchange among the clusters in the current work, as well as a greater fraction of monomer in the CSDs. Both of these observations are in keeping with the shorter (though broader) repulsive hump seen in the pore-forming potentials. In prior work, due to the additive effects of the relatively sharp and tall repulsive humps of the constituent particles, each cluster acted essentially as a re-normalized object, and so the CSDs were only subtly sensitive to changes in density. Here however, the cluster size is much more responsive to changes in overall density.

\clearpage

\subsection*{Visualization of the Relative Entropy Potential Simulations}
\begin{figure}[htb!]
  \includegraphics{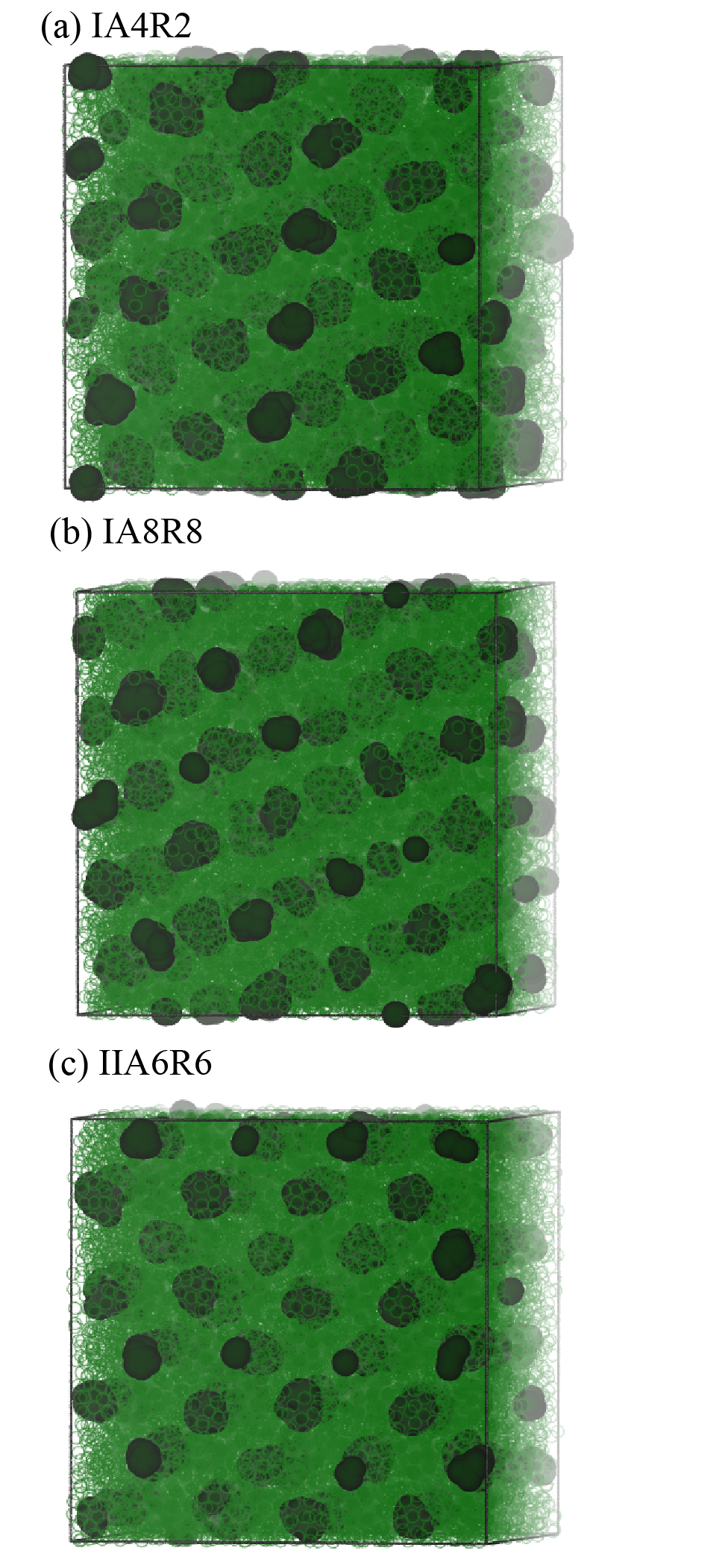}
  \caption{A visualization of the particles and void space as in Fig. 1c of the main text, where the former are lighter green and transparent and the latter is darker and opaque. Ordered, discrete pores are observed in all cases.}
  \label{fgr:FigureRelEntVis}
\end{figure}


%


\end{document}